\documentclass[aps, prl, twocolumn, superscriptaddress, nofootinbib, amsmath, amssymb]{revtex4-2}
\usepackage[utf8]{inputenc}
\usepackage{graphicx}  
\usepackage{mathrsfs}
\usepackage{placeins}
\usepackage{afterpage}
\usepackage{amsthm}
\usepackage{bbold} 
\usepackage{braket}
\usepackage{physics}
\usepackage{calc}
\usepackage[normalem]{ulem}
\usepackage[breaklinks=true,colorlinks,citecolor=blue,linkcolor=blue,urlcolor=blue]{hyperref}
\usepackage{subcaption}
\usepackage{bigints}
\usepackage{hhline}
\usepackage{multirow}
\usepackage{multirow}
\usepackage{rotating}

\usepackage{xargs} 

\usepackage{comment}

\usepackage[font=small, labelfont=bf, textfont={small,it}]{caption}
\makeatletter
\newcommand{\fmarki}{*}
\newcommand{\fmarkii}{\ensuremath{\dagger}}
\newcommand{\fmarkiii}{\ensuremath{\ddagger}}
\newcommand{\fmarkiv}{\ensuremath{\mathsection}}
\newcommand{\fmarkv}{\ensuremath{\mathparagraph}}
\newcommand{\fmarkvi}{\ensuremath{\|}}
\newcommand{\fmarkvii}{**}
\newcommand{\fmarkviii}{\ensuremath{\dagger\dagger}}
\newcommand{\fmarkix}{\ensuremath{\ddagger\ddagger}}
                
\def\@fnsymbol#1{{\ifcase#1\or \fmarki\or \fmarkii\or \fmarkiii\or \fmarkiv\or \fmarkv\or \fmarkvi\or \fmarkvii\or \fmarkviii\or \fmarkix \else\@ctrerr\fi}}
\makeatother

\renewcommand{\fmarkix}{\ensuremath\ddag\ddag}
\def\@fnsymbol#1{{\ifcase#1\or \fmarki\or \fmarkii\or \fmarkiii\or \fmarkiv\or \fmarkv\or \fmarkvi\or \fmarkvii\or \fmarkviii\or \fmarkix \else\@ctrerr\fi}}
\makeatother

\begin{document}

\title{Yang--Mills topology on four-dimensional triangulations}

\author{Giuseppe~Clemente}
\email{giuseppe.clemente@unipi.it}
\affiliation{Dipartimento di Fisica dell'Universit\`a di Pisa and INFN --- Sezione di Pisa, Largo Pontecorvo 3, I-56127 Pisa, Italy.}

\author{Massimo~D'Elia}
\email{massimo.delia@unipi.it}
\affiliation{Dipartimento di Fisica dell'Universit\`a di Pisa and INFN --- Sezione di Pisa, Largo Pontecorvo 3, I-56127 Pisa, Italy.}

\author{Dániel~Németh}
\email{daniel.nemeth@ru.nl}
\affiliation{Institute for Mathematics, Astrophysics and Particle Physics, Radboud University, Heyendaalseweg 135, 6525 AJ Nijmegen, The Netherlands}
\affiliation{{\small{Institute of Theoretical Physics, Jagiellonian University, \L ojasiewicza 11, Kraków, PL 30-348, Poland.}}}

\author{Gianmarco~Simonetti}
\email{g.simonetti8@studenti.unipi.it}
\affiliation{Dipartimento di Fisica dell'Universit\`a di Pisa and INFN --- Sezione di Pisa, Largo Pontecorvo 3, I-56127 Pisa, Italy.}

\begin{abstract}
	We consider 4D $SU(N)$ gauge theories coupled to gravity in the Causal Dynamical Triangulations (CDT) approach,
	focusing on the topological classification of the gauge path integral over fixed triangulations. We discretize the topological charge
	and, after checking the emergence of topology and the continuum scaling on flat triangulations,
	we show that topology emerges on thermalized triangulations only in the so-called de Sitter phase of CDT, thus enforcing the link
	between such phase and semiclassical spacetime. We also provide a tool to visualize the topological structures.
\end{abstract}

\maketitle

\textit{Introduction} --- Triangulations are useful lattice discretizations of curved spacetime~\cite{Regge:1961px},
represented as the gluing of elementary flat regions
known as simplices, like triangles in 2D, tetrahedra in 3D, and so on. They form the basis for the approach to Quantum Gravity (QG), known as Dynamical Triangulations (DT),  consisting in the representation of the regularized Euclidean QG path integral in terms of such configurations, then looking for critical points
within the space of bare parameters as a possible location for
renormalized theory.

The version in which causality is enforced by requiring a well-defined time foliation
(Causal Dynamical Triangulations, or CDT ~\cite{Ambjorn:2012jv,Loll:2019rdj}) is revealing particularly promising in 4D,  because of a nontrivial phase diagram with critical surfaces at the boundary of regions where
semiclassical features emerge.
Within this context, the study of quantum field theories (QFT) coupled to triangulations is of utmost importance. First, it provides an effective tool for
studying QFT on curved spacetime, which is interesting by itself. Second, the path toward renormalized QG should go, sooner or later,
through a full formulation of the theory coupled to other Standard Model fields. In this respect, gauge field theories are a primary goal.

While the program of coupling gauge fields to CDT has been fully accomplished in 2D~\cite{Ambjorn:1999ix,Ambjorn:2013rma,Candido:2020ybd},
its extension to 4D is not easy, at least when building a Markov chain to explore the configuration space of the full
theory~\cite{Clemente:2023sft}. In this context, the present study is an intermediate step, yet addressing
an important aspect.

We investigate $SU(N)$ gauge theories discretized on quenched triangulations, meaning that the spacetime geometry stays
fixed while gauge fields evolve, although we consider different triangulations, going from quasi-flat\footnote{Unlike the
	2D case, it is not possible to generate a flat triangulation of a manifold via equilateral simplices, so we employ quasi-flat
	triangulations, with minimal deviations from being flat.}
to those sampled from the CDT path integral.
Among the possible gauge invariant observables, we focus on gauge topology.

There are at least two good reasons for that. First, the classification of the path integral
into topological sectors originates many nonperturbative features of 4D $\mathrm{SU}(N)$ gauge theories, like the nontrivial dependence  on the CP-violating
parameter $\theta$. Second, such classification is intimately connected to spacetime structure, since it measures the winding over the gauge group
of gauge fields on the 4D spacetime boundary. For instance, it seems hard to find nontrivial topology on triangulations with effective dimension less than
four, similarly to what happens for gauge theories
at asymptotically high temperatures,
which are effectively three-dimensional and for which indeed topological fluctuations are strongly suppressed~\cite{Gross:1980br}.

Therefore, the study of gauge topology could better characterize the properties of spacetime geometries sampled in CDT simulations, in a novel way not considered before.
From a merely algorithmic point of view, the very definition of topology will require
to fix a global orientation for the triangulation, a problem not afforded in previous studies.
Our program goes through three main steps: {\em (i)} implementation of gauge fields on spacetime triangulations,
{\em (ii)} sampling of gauge fields to reproduce the discretized Yang--Mills path integral
on some selected triangulations, which are either
built \emph{ad hoc} or sampled after CDT thermalization through the Einstein--Hilbert action, i.e., we still neglect
the backreaction of gauge fields on triangulation sampling, and
{\em (iii)} definition and investigation of observables capturing topological properties of gauge fields.

\textit{Implementation and sampling} -- In dimension $d > 2$, the only way to correctly count gauge degrees of freedom is to assign a different local gauge to each
simplex~\cite{Ambjorn:1999ix}.
Then, the elementary gauge variables are the $\mathrm{SU}(N)$ parallel transporters connecting adjacent simplices, i.e.,~living on dual graph links~\cite{Candido:2020ybd, Clemente:2023sft}.

The discretization of gauge invariant observables is not unique, however, it relies in general on the definition
of traces of products of gauge variables around closed loops. The \textit{plaquette} $\Pi_b$ is the most elementary
possibility, defined as a product of dual-link variables around a bone $b$ of the triangulation (a $(d-2)$-simplex, i.e., a triangle in $d = 4$).
In the naive continuum limit, similarly to what happens on standard hypercubic lattices,
it has the following correspondence with the field strength tensor $F_{\mu\nu}$, where $\mu\nu$ are directions
orthogonal to the bone $b$
\begin{equation}
	\Pi_b \simeq \exp\left(i g n_b \mathcal{A} F_{\mu\nu}\right),
\end{equation}
where the elementary plaquette area contribution $\mathcal{A} n_b$ is given as the product of the elementary area $\mathcal{A}$ and the number
of simplices around the bone $n_b$, while $g$ denotes the bare gauge coupling constant.

The second-order expansion of $Tr (\Pi_b)$ gives, apart from constant additive factors, $- N n_b^2 g^2 \mathcal{A}^2 Tr (F_{\mu\nu}^2) / 2$, so one can define
\begin{equation}\label{eq:un-YMact_cdt}
	S_{\text{YM}} \equiv - \beta\!\!\!\! \sum\limits_{b \in \mathcal{T}^{(d-2)}}\!\! \frac{\widetilde{\Pi}_b}{n_b} \, , \quad
	\mbox{$\widetilde{\Pi}_b \equiv \lbrack \frac{1}{N} Re Tr \Pi_b - 1 \rbrack$}
\end{equation}
The factor $n_b \mathcal{A}^2$ counts the spacetime volume pertaining to each $b$, apart from a missing factor $\sqrt{5}$, so the whole sum returns, in the naive continuum limit, the volume integral of  $Tr (F_{\mu\nu}^2)$, averaged over all possible directions, times a factor $g^2/(2 N \sqrt{5})$. The usual continuum expression of the Yang--Mills action coincides with this form, apart from a factor 6, coming from the sum over all orthogonal direction pairs. That finally fixes the relation between $g$ and the inverse gauge coupling $\beta$,
\begin{equation}
	\beta = \frac{12 \sqrt{5}\, N}{g^2} \, .
	\label{eq:beta_def}
\end{equation}
The sampling of gauge configurations according to $\exp(-S_{YM})$ can go through usual pure gauge algorithms:
we adopted a standard heat-bath algorithm~\cite{Creutz:1980zw, Cabibbo:1982zn}.
For quasi-flat triangulations, the different spacetime regularization
keeps the asymptotic behavior of the lattice spacing $a(\beta)$ unchanged up to two loops:
\begin{equation}
	\bar a \equiv \Lambda_L a \simeq (\beta_0 g^2)^{\beta_1/(2 \beta_0^2)} e^{-1/(2 \beta_0 g^2)}
	\label{eq:2loopbeta}
\end{equation}
where $3 (4 \pi)^2 \beta_0 = 11 N$ and $3 (4 \pi)^4 \beta_1 = 34 N^2$ for pure Yang--Mills. Of course both $\Lambda_L$ and the regime where such asymptotic behavior sets in are \emph{a priori} unknown.
The issue of continuum limit on thermalized triangulations is far less trivial, because
of the intrinsic length scales characterizing each triangulation,
and could be properly afforded only in a fully dynamical context.

\textit{The observable} ---
As for the definition of topological charge, its continuum expression in 4D is the following:
\begin{equation}\label{eq:topcharge_cont}
	Q = \frac{g^2}{32 \pi^2} \bigintssss\!\! dx\; \varepsilon_{\mu \nu \alpha \beta} \Tr \Big[ F^{\mu \nu}(x) F^{\alpha \beta}(x)\Big].
\end{equation}
We can expand the expression for the field strength tensor around the center of each simplex (i.e., a node of the dual graph) in terms of discretized variables in a local chart. Let us consider a specific simplex $\sigma$ and an overcomplete set of unit vectors ${\{\vec{e}_{A,\sigma}\}}_{A=1,\dots,5}$, directed from the center of $\sigma$ to each of its five neighbors––these vectors are related to a flat Cartesian basis at $\sigma$ via a change of basis,  represented by the matrix $e^\mu_{A,\sigma}$. We can then write
\begin{align}\label{eq:Fmunu_Fbar_rel}
	F^{\mu \nu}_\sigma = e^\mu_{A,\sigma} e^\nu_{B,\sigma} \bar{F}^{AB}_\sigma,
\end{align}
where $\bar{F}^{AB}_\sigma$ can now be related to the plaquette built by looping around the
bone $b_{(AB\sigma)}$ identified by the neighbors $A$ and $B$:
\begin{equation}\label{eq:plaq_eq}
	\Pi_{b_{(AB\sigma)}} \simeq e^{i g \mathcal{A} n_{b_{(AB\sigma)}} \bar{F}^{AB}_\sigma}
	\simeq 1+ig \mathcal{A} n_{b_{(AB\sigma)}} \bar{F}^{AB}_\sigma.
\end{equation}

From Eq.~\eqref{eq:plaq_eq} we can read off,
at the leading order in $a$
\begin{equation}
	\bar{F}^{AB}_\sigma \simeq \frac{1}{g \mathcal{A} n_{b_{(AB\sigma)}}} \mathfrak{Im}\big[ \Pi_{b_{(AB\sigma)}}\big]
	\, ; \
	\mathfrak{Im}\big[\Pi\big] \equiv \frac{\Pi -\Pi^\dagger}{2i}.
\end{equation}
Substituting into the expression for the topological charge density  (Eq.~\eqref{eq:topcharge_cont} before integration), we obtain
\begin{equation}\label{eq:topchargedens_discr}
	\rho_\sigma \simeq \rho^{ABMN}_{\sigma} \bar{\varepsilon}_{ABMN,\sigma}
\end{equation}
where we defined the charge four-form density as
\begin{equation}
	\rho^{ABMN}_{\sigma} \equiv \frac{1}{32 \pi^2 \mathcal{A}^2}  \text{Tr} \Big\{\mathfrak{Im}\Big[ \frac{\Pi_{b_{(AB\sigma)}}}{n_{b_{(AB\sigma)}}}\Big] \mathfrak{Im}\Big[ \frac{\Pi_{b_{(MN\sigma)}}}{n_{b_{(MN\sigma)}}}\Big]\Big\},
\end{equation}
and the Levi--Civita symbol as
\begin{equation}
	\bar{\varepsilon}_{ABMN,\sigma} \equiv \varepsilon_{\mu \nu \alpha \beta} e^\mu_{A,\sigma} e^\nu_{B,\sigma} e^\alpha_{M,\sigma} e^\beta_{N,\sigma}
\end{equation}
expressed in terms of a complete set of basis vectors.
Finally, from Eq.~\eqref{eq:topchargedens_discr}
one can get the topological charge $Q_L$ by integrating $\rho^{ABMN}_{\sigma}$
over spacetime consisting of $N_4$ number of simplices with individual volume
$V_{\sigma} = \sqrt{5}/96$ resulting in $V = N_4 V_\sigma a^4$ 4D volume.

An additional factor $1/5$ stems from the five possible choices of the four independent directions used to build the components of the field strength tensor. Notice that this construction is very similar to the standard symmetrized clover discretization adopted on flat hypercubic lattices, where the number of links coming out from each site is 8, instead of 5 as in our case. Explicitly, $Q_L$ reads
\begin{equation}\label{eq:topcharge_def}
	Q_L=\frac{1}{5}\sum_{\sigma\,\in\,\mathcal{T}}\rho_{\sigma}V_\sigma .
\end{equation}
Finally, a consistent ordering of the vectors $\vec{e}_{A,\sigma}$ over each simplex must be implemented, according to the global orientation of $\mathcal{T}$,
in order to sum the different contributions coherently. See Appendix A for more details.

Topology is well defined only in the continuum limit. A standard procedure, adopted on regular lattices to recover an approximate topological classification, is to iteratively smooth gauge configurations so as to remove ultraviolet noise, typically by minimizing the gauge action and exploiting the fact that topological sectors are characterized by local minima, corresponding to \mbox{(anti-)}instantons in the continuum. We will follow this approach, exploiting in particular the so-called cooling algorithm~\cite{Berg:1981nw, Iwasaki:1983bv, Itoh:1984pr, Teper:1985rb, Ilgenfritz:1985dz, Campostrini:1989dh, Alles:2000sc, Bonati:2014tqa}.
We stress that, as for the discretization of other gauge invariant observables, the definition of the cooling procedure is not unique.

\textit{Numerical Results} --
We start from quasi-flat triangulations of a toroidal hypercube~\cite{SCOTTMARA1976170, Ambjorn:2016fbd},
since this important benchmark should approximate results obtained on standard hypercubic lattices.
We will mainly illustrate results for $SU(3)$, with similar findings obtained for $SU(2)$.

\begin{figure}[t!]
	\centering
	\includegraphics[width=0.5\textwidth]{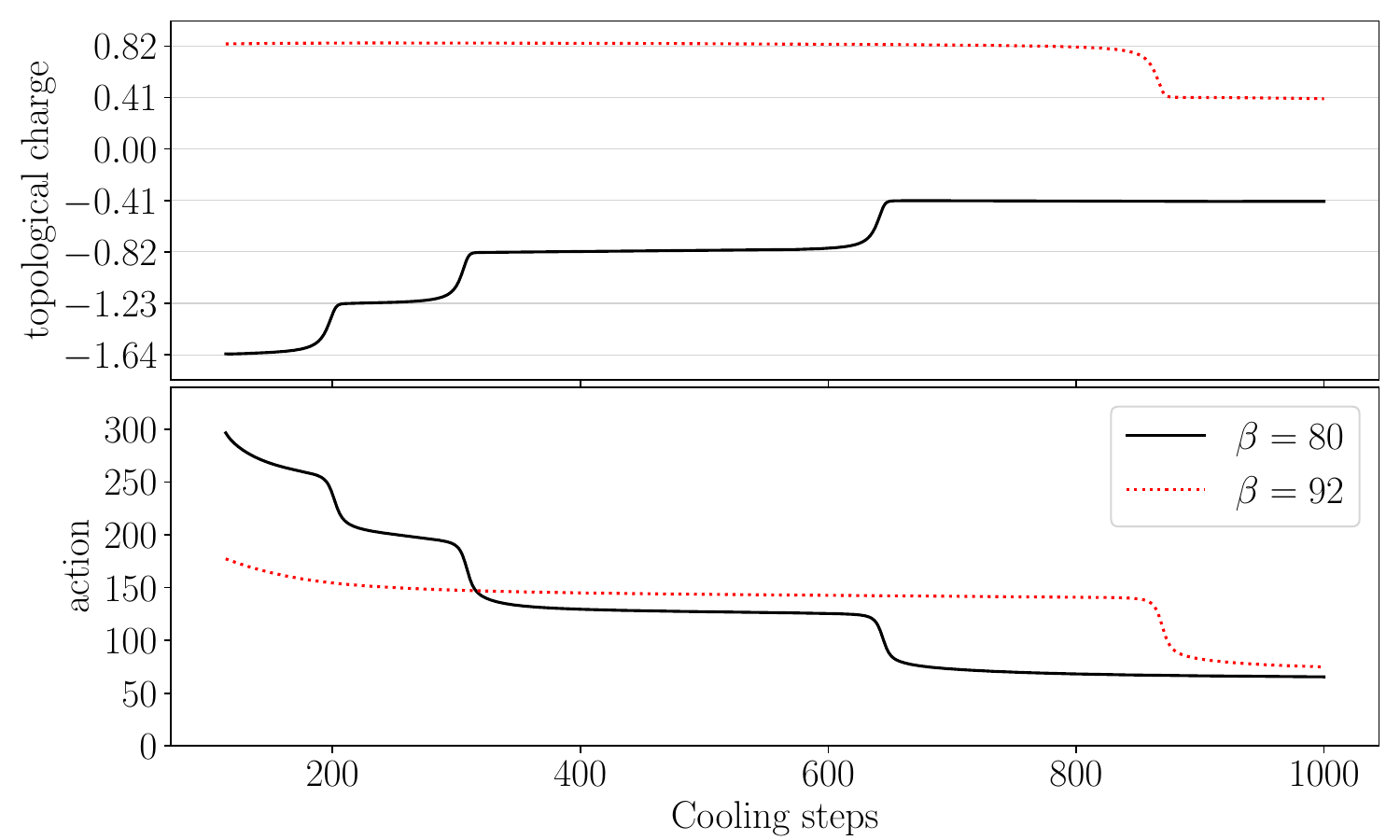}
	\caption{$Q_L$ and $S_{YM}$ evolution during cooling for two $SU(3)$ configurations sampled at different $\beta$ values on quasi-flat triangulations with $T = 40$.}
	\label{fig:descent}
\end{figure}

\begin{figure}[t!]
	\centering
	\includegraphics[width=0.48\textwidth]{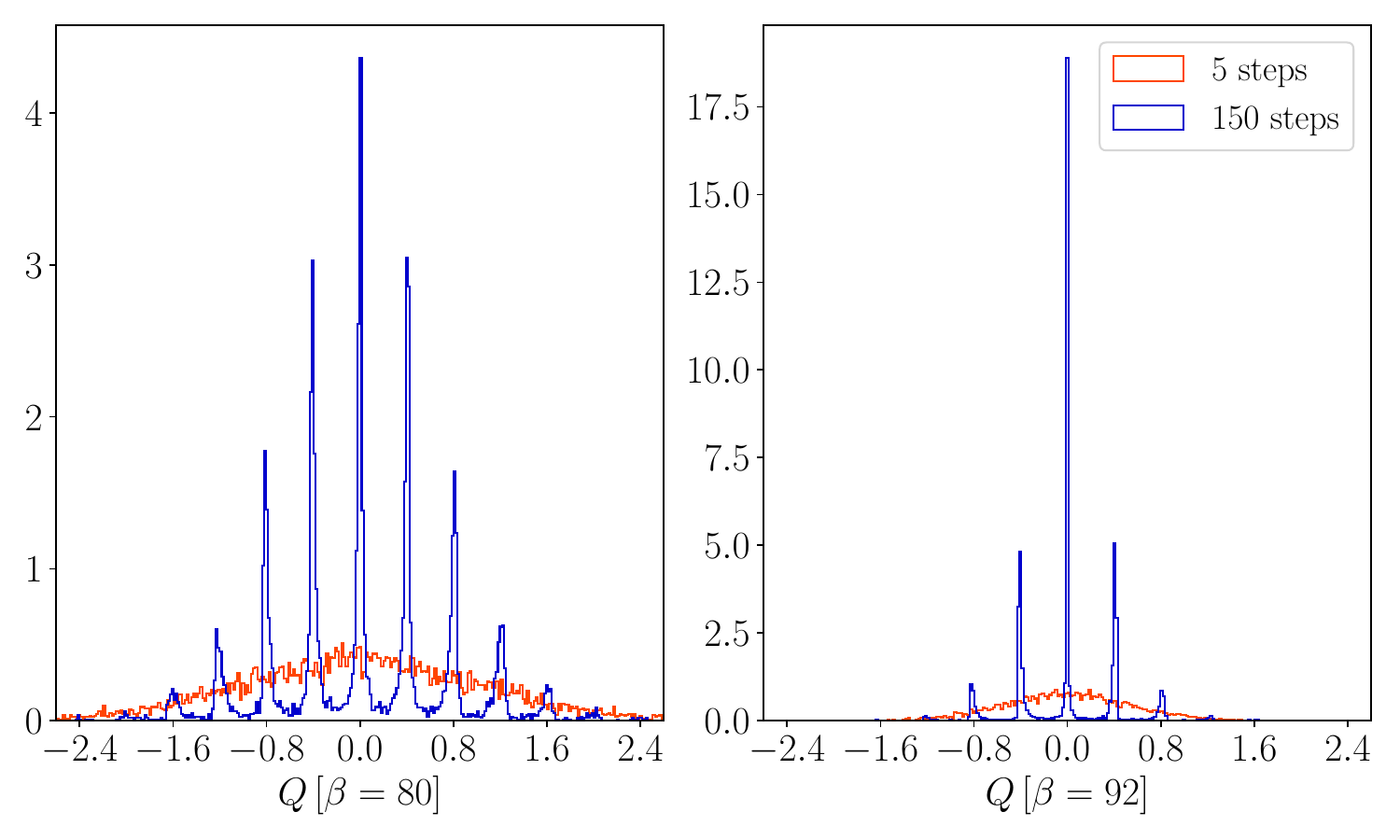}
	\caption{Histograms of $Q_L$ for $SU(3)$ on quasi-flat $T = 40 $ triangulations after $5$~(red) and $150$~(blue) cooling steps.}
	\label{fig:qhist_init}
\end{figure}

Figure~\ref{fig:descent} shows the evolution of $Q_L$ and $S_{YM}$ during the action minimization for two sampled configurations:
similarly to standard lattice studies, we observe intervals where both quantities experience plateaus,
hinting at the presence of metastable states, approximately dividing the path integral into homotopy classes characterized
by different values of $Q_L$.
The histograms of $Q_L$ taken after a sufficient number of cooling steps (Fig.~\ref{fig:qhist_init}) reveal that
the topological charge distribution peaks around integer multiples of an elementary unit $Q_0$, with $Q_0 \simeq 0.41$ independently of $\beta$.

In principle, one would expect $Q_0 \simeq 1$, however also on standard regular lattices discretization effects make $Q_0 < 1$:
here such effects could be larger, because of the approximate flatness of the triangulation. However, a well-defined topological sector $Q$
can be assigned as follows~\cite{DelDebbio:2002xa, Bonati:2015sqt}
\begin{equation}
	Q = \textrm{round}\{Q_L/Q_0\}
\end{equation}
where “round" denotes the approximation to the closest integer and $Q_0$ is fixed by minimizing
\begin{equation}
	\langle \left( Q_L/Q_0 - \textrm{round}\{Q_L/Q_0\}\right)^2 \rangle \, .
\end{equation}

The distribution of $Q$ is wider at lower $\beta$, which is expected by the running of the lattice spacing with $\beta$.
The variance of $Q$ is proportional to the topological susceptibility $\chi \equiv \langle Q^2 \rangle / V$,
therefore, with $\chi$ being a physical quantity, $\langle Q^2 \rangle$ is expected to run as $a^4$ at fixed triangulations.
That gives us the possibility to check if the asymptotic continuum scaling of $a$, Eq.~\eqref{eq:2loopbeta}, is reached within the explored range of $\beta$.

\begin{figure}[t!]
	\centering
	\includegraphics[width=0.5\textwidth,trim={0cm 0.4cm 0 0cm},clip]{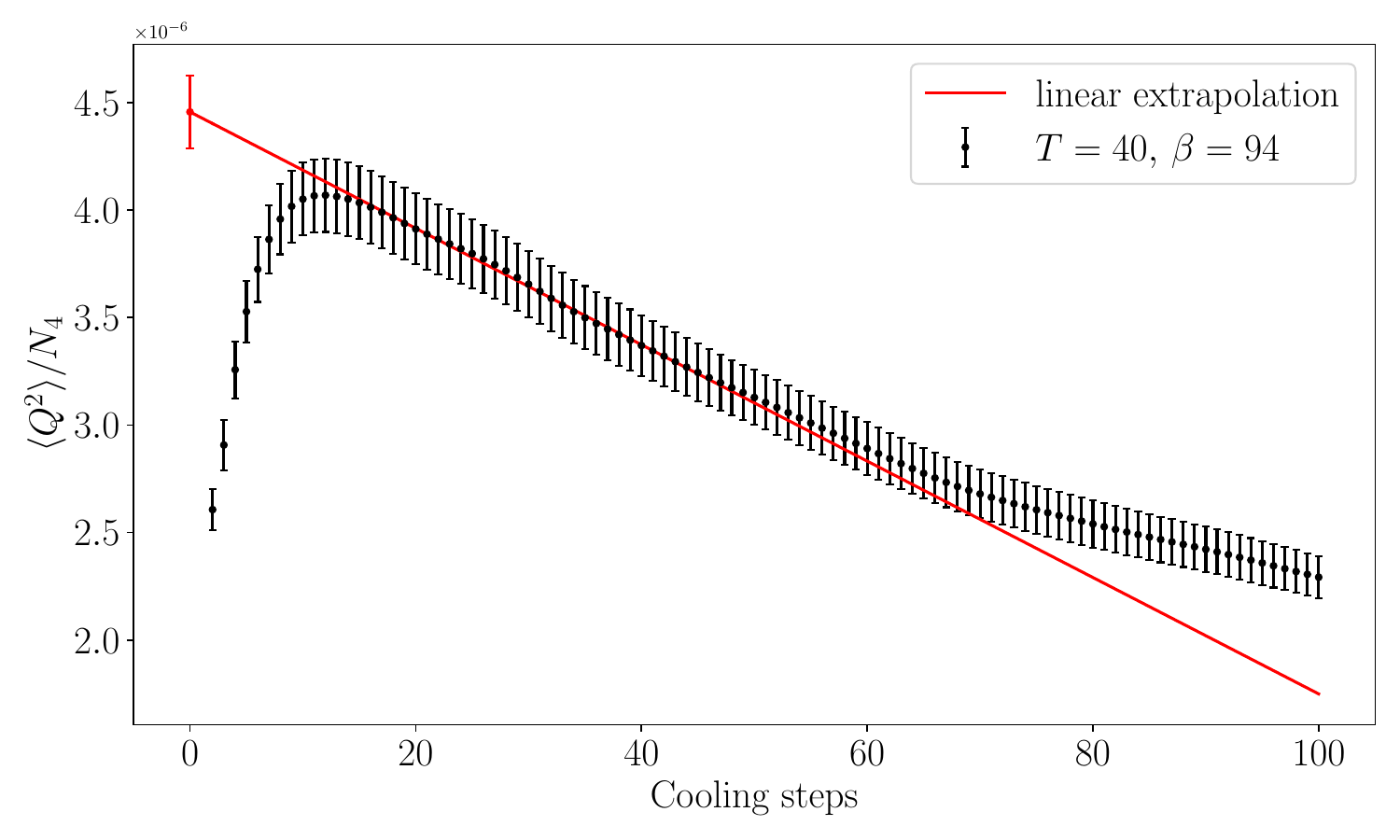}
	\includegraphics[width=0.5\textwidth,trim={0cm 0.5cm 0 0.2cm},clip]{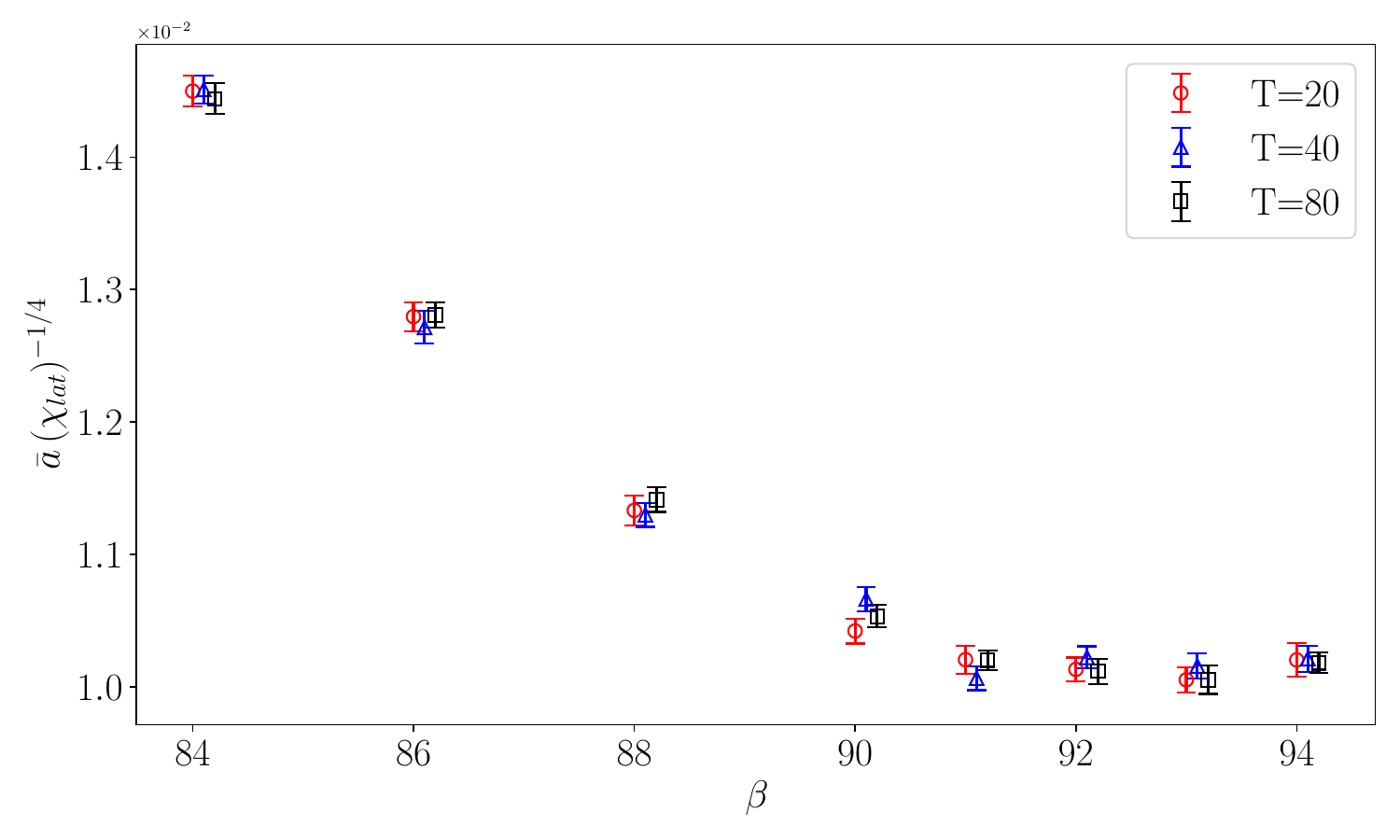}
	\caption{Topological susceptibility for $SU(3)$ gauge fields in a quasi-flat toroidal triangulation.}
	\label{fig:toposusc}
\end{figure}

To this end, we compute the dimensionless susceptibility $\chi_{lat} \equiv \langle Q^2 \rangle / N_4$ for each dataset.
Since topological configurations are only metastable, the smoothing procedure, apart from cleaning the topological signal,
leads also to a partial loss of it. For this reason, it is usual to perform an extrapolation to zero smoothing,
choosing a suitable range of cooling steps for which unphysical fluctuations have already been killed,
while the unwanted loss of signal proceeds linearly with the smoothing.
An example is shown on the upper side of Fig.~\ref{fig:toposusc}.

The so-extrapolated quantity is shown in the lower part of the same figure,
where we report in particular $\bar a\ \chi_{lat}^{-1/4}$ as a function of $\beta$
for different values of the temporal extension $T$,
where $\bar a$ is the 2-loop $\beta$ function in Eq.~\eqref{eq:2loopbeta}. This combination should approach a constant value in the small coupling regime where Eq.~\eqref{eq:2loopbeta} becomes a good approximation. Our results show that, luckily enough, we have at least a small window where this is true. Moreover, the absence of $T$ dependence, within errors, shows that also finite size effects should be reasonably under control.

Assuming the standard pure gauge value from the Witten--Veneziano mechanism~\cite{Witten:1979vv,Veneziano:1979ec},
$\chi^{1/4} \sim 180$~MeV, the measured value of $\chi_{lat}$ gives us an estimate of $a \simeq (\chi_{lat}/(\chi V_{\sigma}))^{1/4}$.
We get $a \sim 0.2$~fm, or in terms of the dual lattice spacing $a' = a / \sqrt{10} \sim 0.06$~fm,
at the border of the explored region, $\beta = 94$.
It is interesting to notice that this is in the same ballpark of lattice spacings where scaling sets in for standard lattices.
Moreover, that tells us \emph{a posteriori} that the spacetime extensions are at least $O(1~{\rm fm})$ for all explored values of $\beta$,
which is consistent, at least for pure gauge theories, with the absence of significant finite size effects.


Another interesting question is how the topological charge density is distributed over the triangulated spacetime.
Standard lattice simulations have shown that the original distribution, i.e., before prolongated smoothing is performed, differs
substantially from a simple instanton picture, with a single bump for each winding unit, corresponding instead to a nontrivial delocalized structure, characterized by many local clusters (see, e.g., Refs.~\cite{Buividovich:2011cv, Ilgenfritz:2007xu, Alexandru:2005bn}).
In order to investigate this aspect, we have developed a visualization tool, which is based on the definition
of pseudo-Cartesian coordinates on the triangulation~\cite{Ambjorn:2021ubd,Ambjorn:2021fkp,Ambjorn:2021uge}.
Details are reported in Appendix B and fully confirm the standard lattice picture described above.

Let us now discuss the interesting case of thermalized configurations.
The observability of gauge topology is tied to the universality class and effective dimension of the underlying geometry.
The phase diagram of CDT~\cite{Ambjorn:2019lrm, Ambjorn:2018qbf} is characterized by two bare parameters $k_0$ and $\Delta$, which are related to the Newton and cosmological constants, as well as to a spacetime anisotropy parameter for the lattice spacings in the causal lattice.
The four geometric phases found include a branched-polymer phase with the universality class of continuum random trees,
two other phases with infinite Hausdorff dimension and a phase representing semiclassical spacetimes~\cite{Ambjorn:2011ph} with emergent effective action called de Sitter phase~\cite{Ambjorn:2008wc}.

\begin{figure}[t!]
	\centering
	\includegraphics[width=0.45\textwidth]{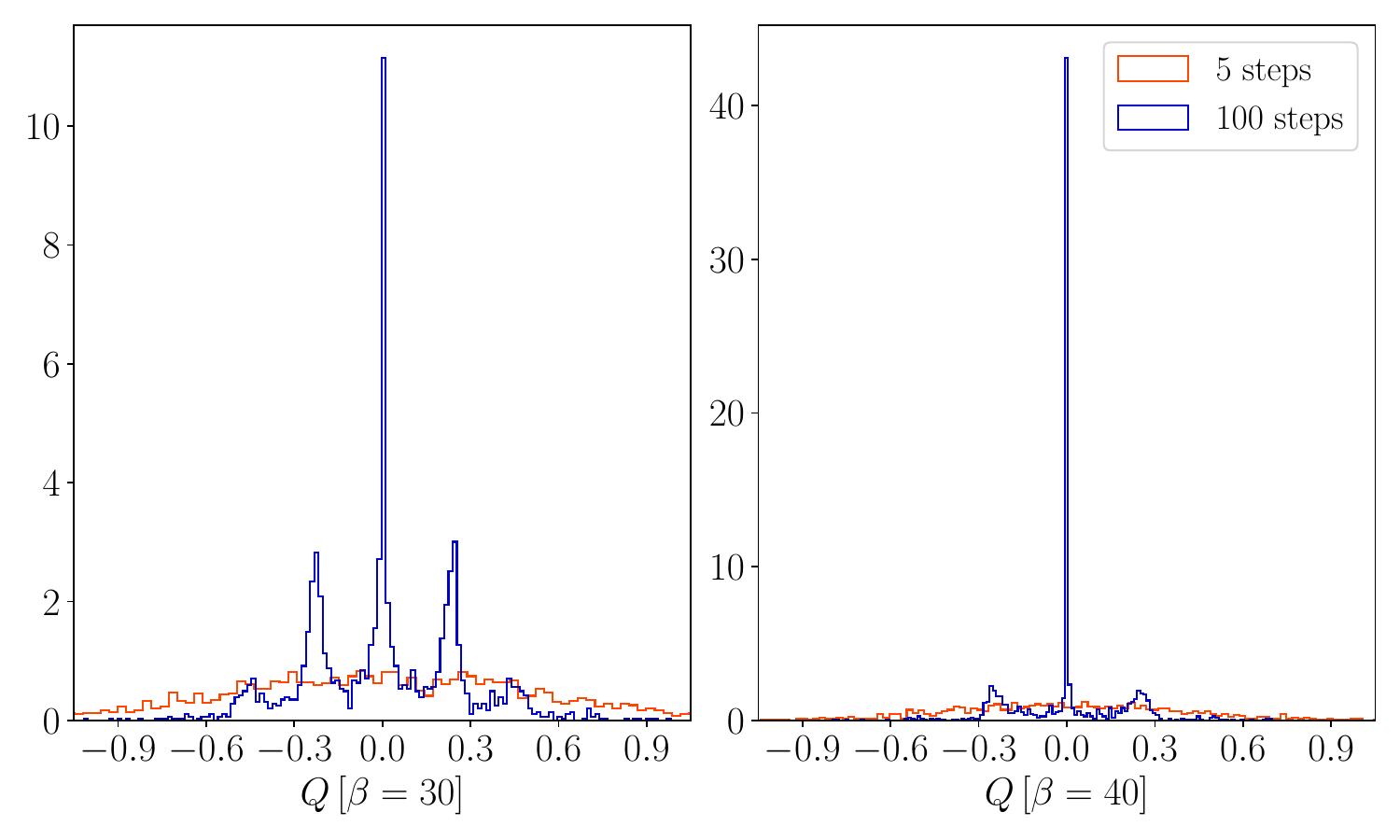}
	\caption{Histograms of $Q_L$ for $SU(3)$ on triangulations thermalized
		in the de Sitter phase, with ${(S^1)}^4$ overall topology, and bare parameters $\kappa_0 = 4.0,\, \Delta = 0$, $T = 20$ and $N_4 = 120k$.}
	\label{fig:qhist_therm}
\end{figure}

Remarkably, we observed nontrivial topological sectors only in the de Sitter phase and, even in this case,
only when the overall numerical triangulation topology was that of a torus $T^4$.
In Fig.~\ref{fig:qhist_therm} we show some examples of the $Q_L$ distribution for this case: after enough smoothing, well-defined peaks appear,
equally spaced by a spacing $Q_0 \sim 0.3$, which differs from the quasi-flat case, probably because of the different
underlying geometries.

\begin{figure}[t!]
	\centering
	\includegraphics[width=0.45\textwidth]{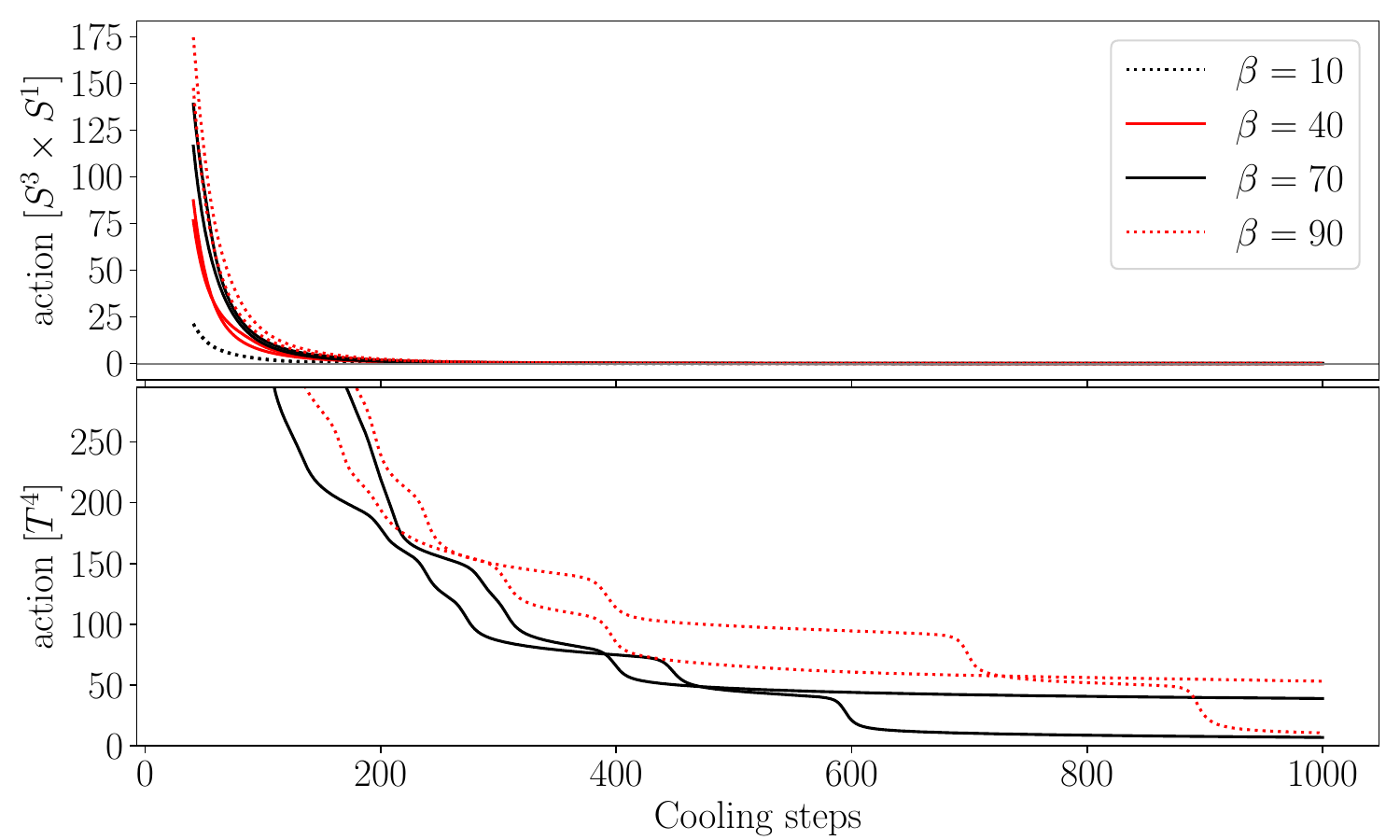}
	\caption{Action descent during smoothing for a few configurations on triangulations thermalized in the de Sitter phase, but with different
		overall topologies.}
	\label{fig:Shist_therm}
\end{figure}

On the contrary, for triangulations thermalized in the de Sitter phase with overall $S^1\cross S^3$ topology, no signs of metastable
structures were observed at all. This is best visible in Fig.~\ref{fig:Shist_therm}, where we compare the gauge action descent during smoothing
for a few configurations in the two cases. The completely different behavior can be hardly interpreted just in terms of the different spacetime
topology, which could influence the overall stability, but not the appearance of metastable gauge configurations. We interpret this as the property
of triangulations being locally mappable to 4D spacetime. Therefore, the different behavior observed for the $S^1\cross S^3$ de Sitter phase
could be due to triangulations in that phase having a local effective dimension different from four: this is a suggestion coming from our
new method to look at these triangulations through the properties of gauge fields, which should be further explored in the future.
We remark that, even if not shown, the action descents observed in all other CDT phases are similar to the upper side of Fig.~\ref{fig:Shist_therm}.

\begin{table}[b!]
	\centering
	\renewcommand\tabcolsep{7pt}
	{\renewcommand\arraystretch{1.1}
		\begin{tabular}{cccc}
			\hline
			\multicolumn{1}{c}{\multirow{2}{*}{$N $}} & \multicolumn{1}{c}{\multirow{2}{*}{$ \beta $}} & \multicolumn{2}{c}{$NS_1/\beta$}                                       \\ \cline{3-4}
			\multicolumn{1}{c}{}                      &                                                & \multicolumn{1}{c}{quasi-flat}   & \multicolumn{1}{c}{de Sitter phase} \\ \hline
			                                          & 20                                             & \multicolumn{1}{c}{$2.30(30)$}   &
			$1.10(20)$                                                                                                                                                          \\
			2                                         & 24                                             &
			\multicolumn{1}{c}{$2.28(25)$}            &
			$1.12(17)$                                                                                                                                                          \\
			                                          & 28                                             &
			\multicolumn{1}{c}{$2.27(21)$}            &
			$1.09(18)$                                                                                                                                                          \\
			\hline
			                                          & 80                                             &
			\multicolumn{1}{c}{$2.28(30)$}            &
			$1.07(23)$                                                                                                                                                          \\
			3                                         & 86                                             &
			\multicolumn{1}{c}{$2.28(28)$}            &
			$1.06(21)$                                                                                                                                                          \\
			                                          & 92                                             &
			\multicolumn{1}{c}{$2.28(26)$}            &
			$1.08(20)$                                                                                                                                                          \\
			\hline
		\end{tabular}}
	\caption{Observed values of the action gap between consecutive topological sectors. The value of $N S_1 / \beta$ expected in the continuum flat case, considering Eq.~\eqref{eq:beta_def}, is about $2.94$.}
	\label{tab:S_1}
\end{table}

Restricting from now on to the de Sitter phase on a torus, we remark that, along the metastable plateaus,
also $S_{YM}$ is approximately a multiple of a well-defined elementary unit $S_1$,
which can be interpreted as the action contribution from a single (anti-)instanton.
In the continuum, the flat case one has~\cite{PhysRevD.14.3432} $S_1 = 8 \pi^2 / g^2$, which,
given our definition of $\beta$ in Eq.~\eqref{eq:beta_def},
means $N S_1 / \beta = 2 \pi^2 / (3 \sqrt{5}) \simeq 2.94$.
In Table~\ref{tab:S_1} we show the estimates of $N S_1 / \beta$ obtained for $N = 2,3$
and for both quasi-flat and thermalized triangulations in the de Sitter phase.
This quantity appears to be quite  stable for different values of $N$ and $\beta$,
but it significantly varies depending on the ensemble of base manifolds.
In particular, the quasi-flat estimate, $N S_1/\beta \simeq 2.3(3)$ is not far from the flat semiclassical expectation in the continuum $2.94$,
while a more significant renormalization appears for thermalized triangulations, where $N S_1/\beta \simeq 1.1(2)$.

In this case, considerations similar to those for the charge quantization
$Q_0$ apply: it is well known that, also on standard regular lattices, the ``instanton'' action is typically smaller than the continuum one,
because of discretization effects (see, e.g., Ref.~\cite{Ilgenfritz:1985dz}).
Moreover, such effects could be enhanced in the case of thermalized configurations,
because of the different underlying spacetime geometry.
Overall, the most important result is the appearance of metastable gauge configurations,
with topological charge and action multiple of well-defined quantities.

\textit{Conclusions} ---
We successfully implemented $SU(N)$ gauge fields on fixed triangulated manifolds,
going from their numerical simulations to the definition and measurement of observables related to gauge topology.
First, we verified the appearance of a topological classification of gauge configurations on quasi-flat triangulations,
checking also the asymptotic scaling to the continuum limit of physical observables such as the topological susceptibility;
and providing a framework to visualize the nontrivial topological structures.

Then, we explored triangulations sampled from the CDT path integral. Remarkably, nontrivial gauge topology appears only in the de Sitter phase
realized on a spacetime torus, thus enforcing the idea that such phase correctly reproduces semiclassical spacetime,
capable of sustaining nontrivial field-theoretic features, such as gauge topology. We have argued that the absence of such features in the corresponding de Sitter phase
realized on $S^1\cross S^3$, as well as in other CDT phases, could reveal different properties of the corresponding triangulations at a local level, such as
an effective local dimension different from four, i.e., the absence of effectively four-dimensional regions capable of sustaining
gauge topology even at a metastable level. We stress that this is one of the main outcomes of our study that should be further explored in the future,
i.e., providing a new tool to explore the geometric properties of triangulations via the nontrivial properties of gauge fields living on them.

Future efforts should also be devoted to the definition of further gauge observables, as well as,
from a long-term perspective, to the numerical simulation of the full gravity+gauge path integral.

\textit{Acknowledgments} --- GC and MD acknowledge support from Fondazione ICSC - National Centre on HPC, Big Data and Quantum Computing - SPOKE 10 (Quantum Computing) and received funding from the European Union Next-GenerationEU - National Recovery and Resilience Plan (NRRP) – MISSION 4 COMPONENT 2, INVESTMENT N. 1.4 – CUP N. I53C22000690001. DN is supported by the VIDI programme with Project No. VI.Vidi.193.048, which is financed by the Dutch Research Council (NWO). The numerical simulations have been performed at the IT Center of the University of Pisa, Jagiellonian University and Radboud University. DN thanks Jakub-Gizbert Studnicki and Andrzej G\"orlich for fruitful discussions and support with the code.

\bibliographystyle{apsrev4-1}
\bibliography{refs}

@article{Regge:1961px,
    author = "Regge, T.",
    title = "{GENERAL RELATIVITY WITHOUT COORDINATES}",
    doi = "10.1007/BF02733251",
    journal = "Nuovo Cim.",
    volume = "19",
    pages = "558--571",
    year = "1961"
}

@article{Ambjorn:2012jv,
    author = "Ambjorn, J. and Goerlich, A. and Jurkiewicz, J. and Loll, R.",
    title = "{Nonperturbative Quantum Gravity}",
    eprint = "1203.3591",
    archivePrefix = "arXiv",
    primaryClass = "hep-th",
    doi = "10.1016/j.physrep.2012.03.007",
    journal = "Phys. Rept.",
    volume = "519",
    pages = "127--210",
    year = "2012"
}

@article{Loll:2019rdj,
    author = "Loll, R.",
    title = "{Quantum Gravity from Causal Dynamical Triangulations: A Review}",
    eprint = "1905.08669",
    archivePrefix = "arXiv",
    primaryClass = "hep-th",
    doi = "10.1088/1361-6382/ab57c7",
    journal = "Class. Quant. Grav.",
    volume = "37",
    number = "1",
    pages = "013002",
    year = "2020"
}

@article{Ambjorn:2016fbd,
    author = {Ambj\o{}rn, J. and Drogosz, Z. and Gizbert-Studnicki, Jakub and G\"orlich, A. and Jurkiewicz, Jerzy and Nemeth, D.},
    title = "{Impact of topology in causal dynamical triangulations quantum gravity}",
    eprint = "1604.08786",
    archivePrefix = "arXiv",
    primaryClass = "hep-th",
    doi = "10.1103/PhysRevD.94.044010",
    journal = "Phys. Rev. D",
    volume = "94",
    number = "4",
    pages = "044010",
    year = "2016"
}

@article{Ambjorn:2021uge,
    author = {Ambj\o{}rn, Jan and Drogosz, Zbigniew and Gizbert-Studnicki, Jakub and G\"orlich, Andrzej and Jurkiewicz, Jerzy and N\'emeth, D. \'aniel},
    title = "{Scalar fields in causal dynamical triangulations}",
    eprint = "2105.10086",
    archivePrefix = "arXiv",
    primaryClass = "gr-qc",
    doi = "10.1088/1361-6382/ac2135",
    journal = "Class. Quant. Grav.",
    volume = "38",
    number = "19",
    pages = "195030",
    year = "2021"
}

@article{Ambjorn:2021fkp,
    author = {Ambj\o{}rn, J. and Drogosz, Z. and Gizbert-Studnicki, J. and G\"orlich, A. and Jurkiewicz, J. and N\'emeth, D.},
    title = "{Matter-Driven Change of Spacetime Topology}",
    eprint = "2103.00198",
    archivePrefix = "arXiv",
    primaryClass = "hep-th",
    doi = "10.1103/PhysRevLett.127.161301",
    journal = "Phys. Rev. Lett.",
    volume = "127",
    number = "16",
    pages = "161301",
    year = "2021"
}

@article{Ambjorn:2021ubd,
    author = {Ambjorn, J. and Drogosz, Z. and Gizbert-Studnicki, J. and Gőrlich, A. and Jurkiewicz, J. and N\'emeth, D.},
    title = "{Cosmic voids and filaments from quantum gravity}",
    eprint = "2101.08617",
    archivePrefix = "arXiv",
    primaryClass = "gr-qc",
    doi = "10.1140/epjc/s10052-021-09468-z",
    journal = "Eur. Phys. J. C",
    volume = "81",
    number = "8",
    pages = "708",
    year = "2021"
}

@article{Ambjorn:2019lrm,
    author = {Ambj\o{}rn, Jan and Gizbert-Studnicki, Jakub and Gőrlich, Andrzej and Jurkiewicz, Jerzy and N\'emeth, D\'aniel},
    title = "{Towards an UV fixed point in CDT gravity}",
    eprint = "1906.04557",
    archivePrefix = "arXiv",
    primaryClass = "hep-th",
    doi = "10.1007/JHEP07(2019)166",
    journal = "JHEP",
    volume = "07",
    pages = "166",
    year = "2019"
}

@article{Ambjorn:2018qbf,
    author = {Ambj\o{}rn, Jan and Gizbert-Studnicki, Jakub and G\"orlich, Andrzej and Jurkiewicz, Jerzy and N\'emeth, D\'aniel},
    title = "{The phase structure of Causal Dynamical Triangulations with toroidal spatial topology}",
    eprint = "1802.10434",
    archivePrefix = "arXiv",
    primaryClass = "hep-th",
    doi = "10.1007/JHEP06(2018)111",
    journal = "JHEP",
    volume = "06",
    pages = "111",
    year = "2018"
}

@article{Ambjorn:2008wc,
    author = "Ambjorn, J. and Gorlich, A. and Jurkiewicz, J. and Loll, R.",
    title = "{The Nonperturbative Quantum de Sitter Universe}",
    eprint = "0807.4481",
    archivePrefix = "arXiv",
    primaryClass = "hep-th",
    reportNumber = "ITP-UU-08-44, SPIN-08-44",
    doi = "10.1103/PhysRevD.78.063544",
    journal = "Phys. Rev. D",
    volume = "78",
    pages = "063544",
    year = "2008"
}

@article{Ambjorn:2011ph,
    author = "Ambjorn, J. and Gorlich, A. and Jurkiewicz, J. and Loll, R. and Gizbert-Studnicki, J. and Trzesniewski, T.",
    title = "{The Semiclassical Limit of Causal Dynamical Triangulations}",
    eprint = "1102.3929",
    archivePrefix = "arXiv",
    primaryClass = "hep-th",
    doi = "10.1016/j.nuclphysb.2011.03.019",
    journal = "Nucl. Phys. B",
    volume = "849",
    pages = "144--165",
    year = "2011"
}

@article{Ambjorn:1999ix,
    author = "Ambjorn, Jan and Anagnostopoulos, K. N. and Jurkiewicz, J.",
    title = "{Abelian gauge fields coupled to simplicial quantum gravity}",
    eprint = "hep-lat/9907027",
    archivePrefix = "arXiv",
    reportNumber = "NBI-HE-99-51, TPJU-7-99",
    doi = "10.1088/1126-6708/1999/08/016",
    journal = "JHEP",
    volume = "08",
    pages = "016",
    year = "1999"
}

@article{Ambjorn:2013rma,
    author = "Ambjorn, J. and Ipsen, A.",
    title = "{Two-dimensional causal dynamical triangulations with gauge fields}",
    eprint = "1305.3148",
    archivePrefix = "arXiv",
    primaryClass = "hep-th",
    doi = "10.1103/PhysRevD.88.067502",
    journal = "Phys. Rev. D",
    volume = "88",
    number = "6",
    pages = "067502",
    year = "2013"
}

@article{SCOTTMARA1976170,
title = {Triangulations for the cube},
journal = {Journal of Combinatorial Theory, Series A},
volume = {20},
number = {2},
pages = {170-177},
year = {1976},
issn = {0097-3165},
doi = {https://doi.org/10.1016/0097-3165(76)90014-5},
url = {https://www.sciencedirect.com/science/article/pii/0097316576900145},
author = {Patrick {Scott Mara}}
}

@article{Candido:2020ybd,
    author = "Candido, Alessandro and Clemente, Giuseppe and D'Elia, Massimo and Rottoli, Federico",
    title = "{Compact gauge fields on Causal Dynamical Triangulations: a 2D case study}",
    eprint = "2010.15714",
    archivePrefix = "arXiv",
    primaryClass = "hep-lat",
    doi = "10.1007/JHEP04(2021)184",
    journal = "JHEP",
    volume = "04",
    pages = "184",
    year = "2021"
}

@Inbook{Clemente:2023sft,
    author="Clemente, Giuseppe
    and D'Elia, Massimo",
    editor="Bambi, Cosimo
    and Modesto, Leonardo
    and Shapiro, Ilya",
    title="{Spectral Observables and Gauge Field Couplings in Causal Dynamical Triangulations}",
    bookTitle="Handbook of Quantum Gravity",
    year="2023",
    publisher="Springer Nature Singapore",
    address="Singapore",
    pages="1--34",
    isbn="978-981-19-3079-9",
    doi="10.1007/978-981-19-3079-9_89-1",
    url="https://doi.org/10.1007/978-981-19-3079-9_89-1"
}

@article{Creutz:1980zw,
    author = "Creutz, M.",
    title = "{Monte Carlo Study of Quantized SU(2) Gauge Theory}",
    doi = "10.1103/PhysRevD.21.2308",
    journal = "Phys. Rev. D",
    volume = "21",
    pages = "2308--2315",
    year = "1980"
}

@article{Cabibbo:1982zn,
    author = "Cabibbo, N. and Marinari, E.",
    title = "{A New Method for Updating SU(N) Matrices in Computer Simulations of Gauge Theories}",
    doi = "10.1016/0370-2693(82)90696-7",
    journal = "Phys. Lett. B",
    volume = "119",
    pages = "387--390",
    year = "1982"
}

@article{PhysRevD.14.3432,
  title = {Computation of the quantum effects due to a four-dimensional pseudoparticle},
  author = {'t Hooft, G.},
  journal = {Phys. Rev. D},
  volume = {14},
  issue = {12},
  pages = {3432--3450},
  numpages = {0},
  year = {1976},
  month = {Dec},
  publisher = {American Physical Society},
  doi = {10.1103/PhysRevD.14.3432},
  url = {https://link.aps.org/doi/10.1103/PhysRevD.14.3432}
}

@article{Berg:1981nw,
    author = "Berg, B.",
    title = "{Dislocations and Topological Background in the Lattice O(3) $\sigma$ Model}",
    reportNumber = "CERN-TH-3076",
    doi = "10.1016/0370-2693(81)90518-9",
    journal = "Phys. Lett. B",
    volume = "104",
    pages = "475--480",
    year = "1981"
}

@article{Iwasaki:1983bv,
    author = "Iwasaki, Y. and Yoshie, T.",
    title = "{Instantons and Topological Charge in Lattice Gauge Theory}",
    reportNumber = "UTHEP-109",
    doi = "10.1016/0370-2693(83)91111-5",
    journal = "Phys. Lett. B",
    volume = "131",
    pages = "159--164",
    year = "1983"
}

@article{Itoh:1984pr,
    author = "Itoh, S. and Iwasaki, Y. and Yoshie, T.",
    title = "{Stability of Instantons on the Lattice and the Renormalized Trajectory}",
    reportNumber = "UTHEP-125",
    doi = "10.1016/0370-2693(84)90609-9",
    journal = "Phys. Lett. B",
    volume = "147",
    pages = "141--144",
    year = "1984"
}

@article{Teper:1985rb,
    author = "Teper, M.",
    title = "{Instantons in the Quantized SU(2) Vacuum: A Lattice Monte Carlo Investigation}",
    reportNumber = "CERN-TH-4208/85",
    doi = "10.1016/0370-2693(85)90939-6",
    journal = "Phys. Lett. B",
    volume = "162",
    pages = "357--362",
    year = "1985"
}

@article{Ilgenfritz:1985dz,
    author = "Ilgenfritz, Ernst-Michael and Laursen, M. L. and Schierholz, G. and Muller-Preussker, M. and Schiller, H.",
    editor = "Loken, S. C.",
    title = "{First Evidence for the Existence of Instantons in the Quantized SU(2) Lattice Vacuum}",
    reportNumber = "DESY-85-108",
    doi = "10.1016/0550-3213(86)90265-8",
    journal = "Nucl. Phys. B",
    volume = "268",
    pages = "693",
    year = "1986"
}

@article{Campostrini:1989dh,
    author = "Campostrini, Massimo and Di Giacomo, Adriano and Panagopoulos, Haralambos and Vicari, Ettore",
    title = "{Topological Charge, Renormalization and Cooling on the Lattice}",
    reportNumber = "IFUP-TH-2/89",
    doi = "10.1016/0550-3213(90)90077-Q",
    journal = "Nucl. Phys. B",
    volume = "329",
    pages = "683--697",
    year = "1990"
}

@article{Alles:2000sc,
    author = "Alles, B. and Cosmai, L. and D'Elia, Massimo and Papa, A.",
    title = "{Topology in 2-D CP**(N-1) models on the lattice: A Critical comparison of different cooling techniques}",
    eprint = "hep-lat/0001027",
    archivePrefix = "arXiv",
    reportNumber = "BICOCCA-FT-99-40, BARI-TH-370-99, IFUP-TH-67-99, UNICAL-TH-99-6",
    doi = "10.1103/PhysRevD.62.094507",
    journal = "Phys. Rev. D",
    volume = "62",
    pages = "094507",
    year = "2000"
}

@article{DelDebbio:2002xa,
    author = "Del Debbio, Luigi and Panagopoulos, Haralambos and Vicari, Ettore",
    title = "{theta dependence of SU(N) gauge theories}",
    eprint = "hep-th/0204125",
    archivePrefix = "arXiv",
    doi = "10.1088/1126-6708/2002/08/044",
    journal = "JHEP",
    volume = "08",
    pages = "044",
    year = "2002"
}

@article{Witten:1979vv,
    author = "Witten, Edward",
    title = "{Current Algebra Theorems for the U(1) Goldstone Boson}",
    reportNumber = "HUTP-79/A014",
    doi = "10.1016/0550-3213(79)90031-2",
    journal = "Nucl. Phys. B",
    volume = "156",
    pages = "269--283",
    year = "1979"
}

@article{Veneziano:1979ec,
    author = "Veneziano, G.",
    title = "{U(1) Without Instantons}",
    reportNumber = "CERN-TH-2651",
    doi = "10.1016/0550-3213(79)90332-8",
    journal = "Nucl. Phys. B",
    volume = "159",
    pages = "213--224",
    year = "1979"
}

@article{Bonati:2015sqt,
    author = "Bonati, Claudio and D'Elia, Massimo and Scapellato, Aurora",
    title = "{$\theta$ dependence in $SU(3)$ Yang-Mills theory from analytic continuation}",
    eprint = "1512.01544",
    archivePrefix = "arXiv",
    primaryClass = "hep-lat",
    reportNumber = "IFUP-TH-2015-14",
    doi = "10.1103/PhysRevD.93.025028",
    journal = "Phys. Rev. D",
    volume = "93",
    number = "2",
    pages = "025028",
    year = "2016"
}

@article{Bonati:2014tqa,
    author = "Bonati, Claudio and D'Elia, Massimo",
    title = "{Comparison of the gradient flow with cooling in $SU(3)$ pure gauge theory}",
    eprint = "1401.2441",
    archivePrefix = "arXiv",
    primaryClass = "hep-lat",
    doi = "10.1103/PhysRevD.89.105005",
    journal = "Phys. Rev. D",
    volume = "89",
    number = "10",
    pages = "105005",
    year = "2014"
}

@article{Gross:1980br,
    author = "Gross, David J. and Pisarski, Robert D. and Yaffe, Laurence G.",
    title = "{QCD and Instantons at Finite Temperature}",
    reportNumber = "PRINT-80-0538 (PRINCETON)",
    doi = "10.1103/RevModPhys.53.43",
    journal = "Rev. Mod. Phys.",
    volume = "53",
    pages = "43",
    year = "1981"
}

@article{Buividovich:2011cv,
    author = "Buividovich, P. V. and Kalaydzhyan, T. and Polikarpov, M. I.",
    title = "{Fractal dimension of the topological charge density distribution in SU(2) lattice gluodynamics}",
    eprint = "1111.6733",
    archivePrefix = "arXiv",
    primaryClass = "hep-lat",
    reportNumber = "ITEP-LAT-2011-12, DESY-11-176",
    doi = "10.1103/PhysRevD.86.074511",
    journal = "Phys. Rev. D",
    volume = "86",
    pages = "074511",
    year = "2012"
}

@article{Ilgenfritz:2007xu,
    author = "Ilgenfritz, E. -M. and Koller, K. and Koma, Y. and Schierholz, G. and Streuer, T. and Weinberg, V.",
    title = "{Exploring the structure of the quenched QCD vacuum with overlap fermions}",
    eprint = "0705.0018",
    archivePrefix = "arXiv",
    primaryClass = "hep-lat",
    reportNumber = "DESY-07-055, HU-EP-07-08, LMU-ASC-79-06, MKPH-T-07-05",
    doi = "10.1103/PhysRevD.76.034506",
    journal = "Phys. Rev. D",
    volume = "76",
    pages = "034506",
    year = "2007"
}

@article{Alexandru:2005bn,
    author = "Alexandru, Andrei and Horvath, Ivan and Zhang, Jian-bo",
    title = "{The Reality of the fundamental topological structure in the QCD vacuum}",
    eprint = "hep-lat/0506018",
    archivePrefix = "arXiv",
    doi = "10.1103/PhysRevD.72.034506",
    journal = "Phys. Rev. D",
    volume = "72",
    pages = "034506",
    year = "2005"
}

\section{Appendix A. Reordering procedure}\label{app:A}
\begin{figure}[h!]
	\centering
	\includegraphics[width=0.4\textwidth]{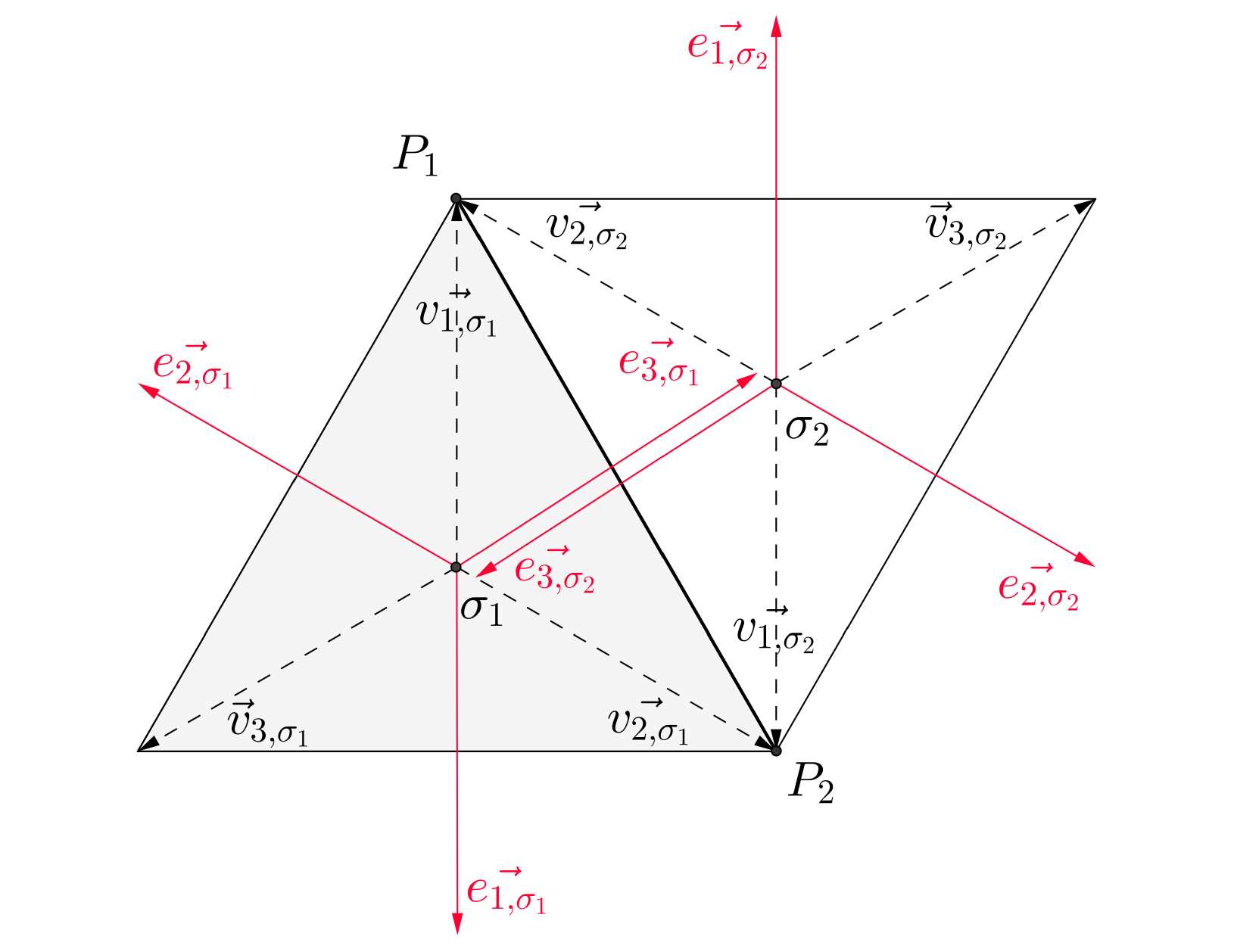}
	\caption{Two-dimensional sketch of the ordering procedure.}
	\label{fig:ordering}
\end{figure}

In order to have a meaningful implementation of the topological charge, the simplices of the triangulation must be properly oriented so that the pseudoscalarity of $Q$ holds. In the particular case of simplicial manifolds, this means that for any pair of adjacent simplices $\sigma_{1,2} \in \mathcal{T}$, the local frames $\{ \vec{e}_{\sigma_{1,2},A}\}$ already mentioned before, must coincide up to a rototranslation,
\begin{equation}\label{eq:same_parity}
	\vec{e}_{A,\sigma_2} = (R_{\sigma_2} \circ t_{\sigma_2 \leftarrow \sigma_1})   \vec{e}_{A,\sigma_1}\quad  \forall A\in\{1,\dots,d\}
\end{equation}
with $R_{\sigma_2} \in SO(d)$ and $t_{\sigma_2 \leftarrow \sigma_1}$, respectively, a rotation defined in the center of $\sigma_2$ and a translation that moves the center of $\sigma_1$ to the center of $\sigma_2$. In other words, Eq.~\eqref{eq:same_parity} just states that the two frames must have the same parity. In the following, we will show a possible strategy to fix a global orientation of $\mathcal{T}$.  This method is simply based on setting the orientation of $\{ \vec{e}_{A,\sigma_2}\}$ assuming that $\{ \vec{e}_{A,\sigma_1}\}$ has already been fixed. To do so, we will make use of the vectors $\{ \vec{v}_{A,\sigma_{1,2}}\}_{A=1,\dots,d}$, which represent the set of vectors connecting the centers of $\sigma_{1,2}$ to their vertices. Let us start by first fixing $\{ \vec{v}_{A,\sigma_{1}}\}$. This can be done, for instance, by symmetrizing and scaling\footnote{We are assuming that $ \{ \vec{e}_{A,\sigma}\} $ are unitary vectors.} $\{ \vec{e}_{A,\sigma_1}\}$ with respect to the center of $\sigma_1$,
\begin{equation}\label{eq:flip}
	\vec{v}_{A,\sigma_1} = -\left\| \vec{v} \right\|  \vec{e}_{A,\sigma_1}\quad \forall A\in\{1,\dots,d\}.
\end{equation}
In this way, the two frames in the expression above have the same parity in even dimensions and opposite parity otherwise. At this point one just needs to fix $\{\vec{v}_{A,\sigma_2}\}$ by requiring that the latter has the same orientation of $\{\vec{v}_{A,\sigma_1}\}$ and use again~\eqref{eq:flip} to retrieve $\{\vec{e}_{A,\sigma_2}\}$ which eventually will satisfy~\eqref{eq:same_parity}. For this purpose, let us label with $\{P_{i}\}_{i=1,\dots,d-1}$ the vertices of the $d-1$-dimensional subsimplex shared by $\sigma_1$, $\sigma_2$: if labels are chosen so that the association between such points and the first $d-1$ vectors of $\{ \vec{v}_{A,\sigma_1} \}$ is
\begin{equation}
	\{ \vec{v}_{A,\sigma_1} \} =  \{ \vec{v}_{1,\sigma_1}(P_1), \dots , \vec{v}_{d-1,\sigma_1}(P_{d-1}),\vec{v}_{d,\sigma_1} \}
\end{equation}
then $\{ \vec{v}_{A,\sigma_2} \} $ can be obtained by exchanging any two directions after that the common vertices have been associated with the vectors of $\sigma_2$ in the same way as for $\sigma_1$. Explicitly, it reads
\begin{equation}
	\{ \vec{v}_{A,\sigma_2} \} =  \tau_{jk} \, \{ \vec{v}_{1,\sigma_2}(P_1), \dots , \vec{v}_{d-1,\sigma_2}(P_{d-1}),\vec{v}_{d,\sigma_2} \},
\end{equation}
where $\tau_{jk}$ is a transpose operator acting as
\begin{equation}
	\begin{split}
		\tau_{jk}&\{\dots \vec{v}_{j,\sigma_2}(P_j),\dots,  \vec{v}_{k,\sigma_2}(P_k),\dots   \}\\ = &\{\dots \vec{v}_{j,\sigma_2}(P_k), \dots, \vec{v}_{k,\sigma_2}(P_j),\dots   \}.
	\end{split}
\end{equation}
Indeed, since the uncommon vertices (one for each simplex) belong to each of the two separated space regions defined by the shared subsimplex, $\{ \vec{v}_{A,\sigma_1} \} $ and  $\{ \vec{v}_{A,\sigma_2} \} $ would have opposite parity if the action of a discrete operator like $\tau$ was neglected. Such a procedure, sketched
in Fig.~\ref{fig:ordering} for the 2D case, can be iterated for all the nearest neighbors of $\sigma_1$ and, therefore, extended to all the simplices of the triangulation. Of course, this can be done consistently, up to pure rotations, only if the given triangulation is globally orientable.

\section{Appendix B. Visualization of nontrivial topological structures}\label{app:B}

By definition, compact triangulations do not admit a natural coordinate system. In~\cite{Ambjorn:2021ubd,Ambjorn:2021fkp,Ambjorn:2021uge},
a method was introduced to define pseudo-Cartesian coordinates in CDT.
A curved Riemannian manifold $M$ can be mapped to another manifold $N$ with a flat metric using a mapping represented by a scalar field $\phi$.
Associating a field with every coordinate direction, the values of the four fields $\phi_{x,y,z,t}$ represent a new field coordinate system.
By choosing a suitable boundary condition, a useful metric structure can be associated with the triangulation,
which can be used to  map distances and structures different from that of the dual lattice.

\begin{figure}[hbt!]
	\centering
	\includegraphics[width=\linewidth]{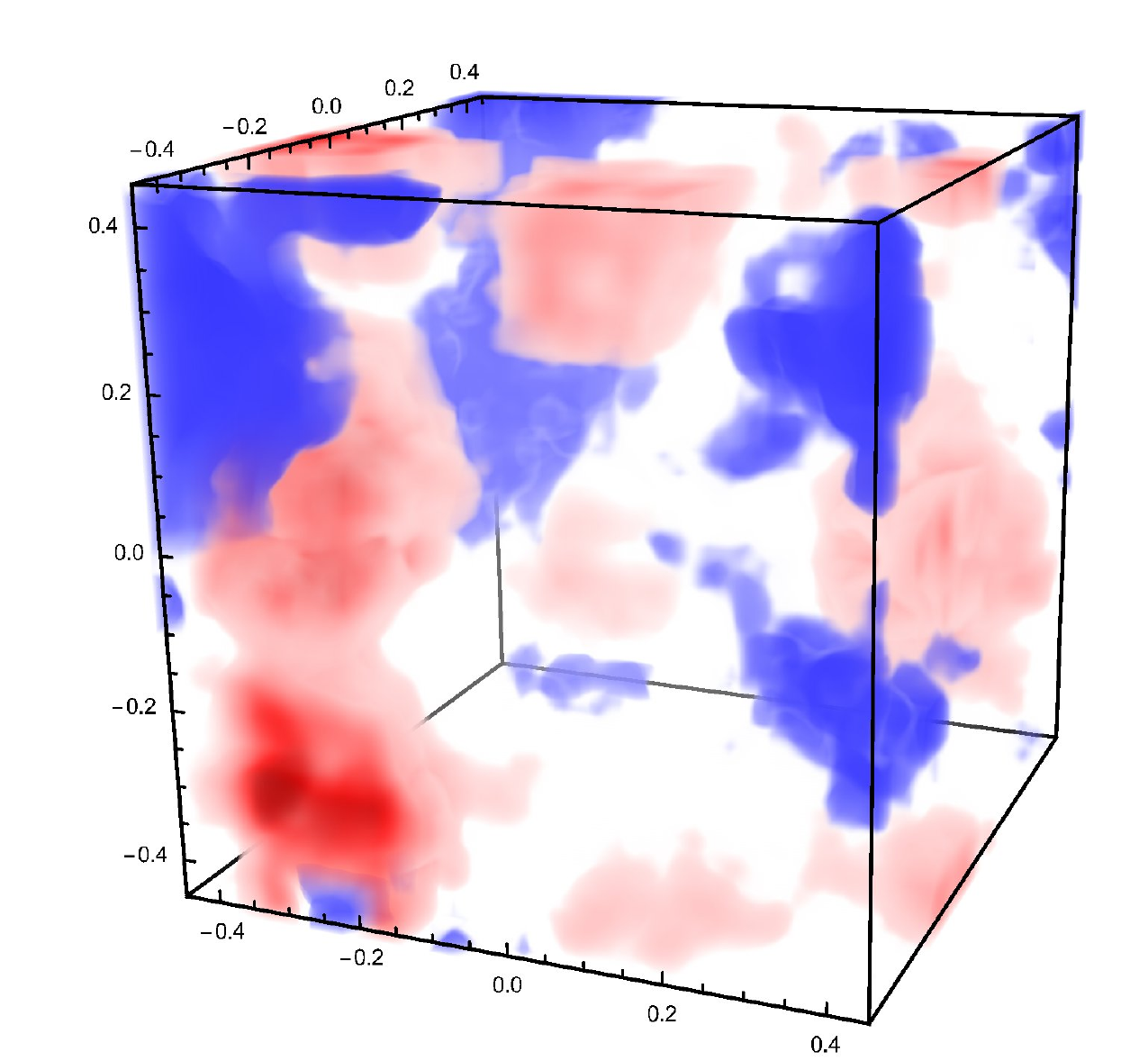}
	\caption{A visualization of the charge density in $\phi_x, \phi_y, \phi_z$
		scalar field coordinate representation after 10 cooling steps for $\beta^{SU(2)} = 30$, using the quasi-flat configuration.
		The topological charge density has been integrated in the time direction.}
	\label{fig:boxes}
\end{figure}

\begin{figure}[hbt!]
	\centering
	\includegraphics[width=\linewidth]{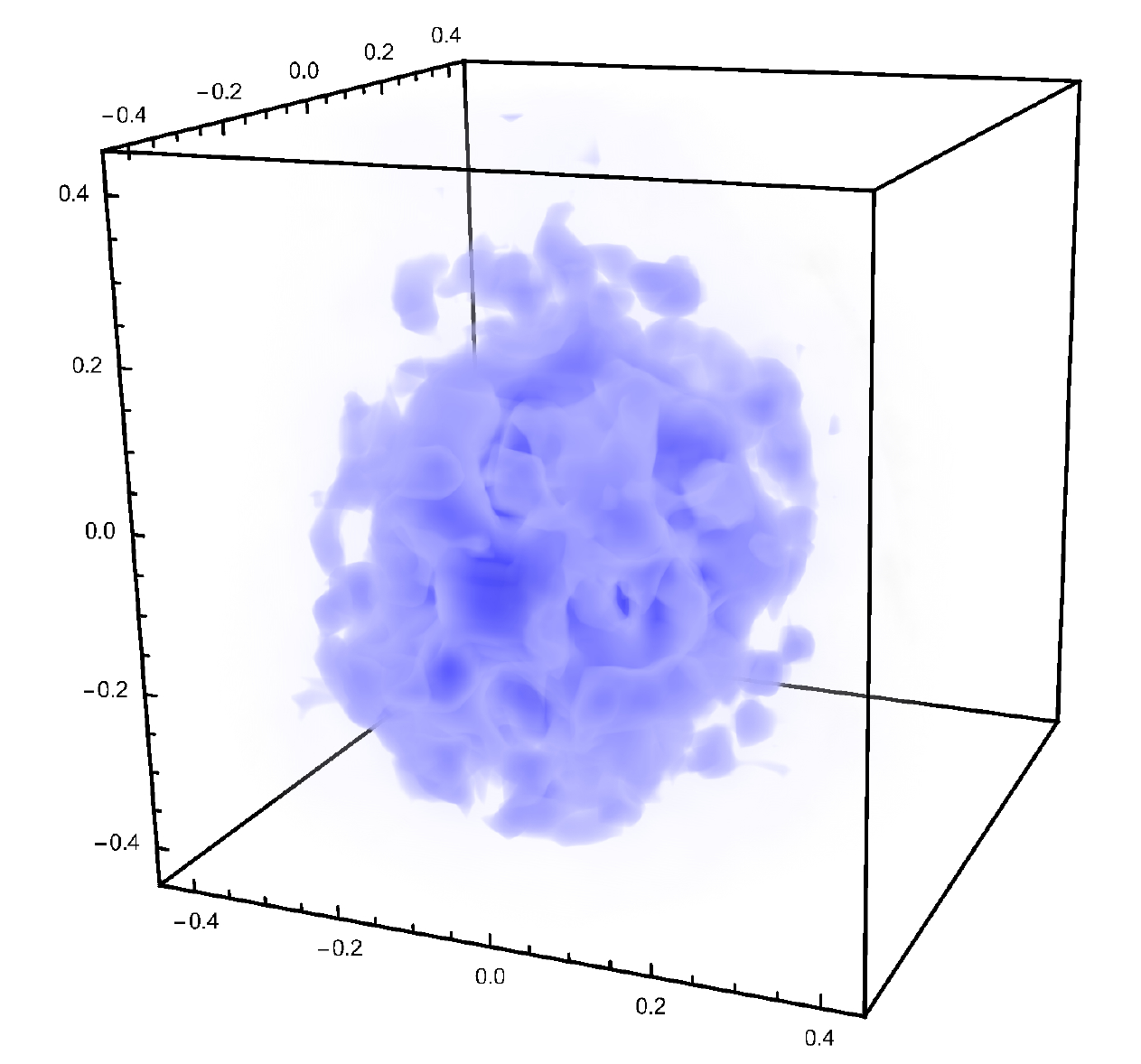}
	\caption{Charge density visualization for the same triangulation and gauge configuration reported in Fig.~\ref{fig:boxes},
		but after prolongated cooling (150 steps). Also in this case, the topological charge density has been integrated in the time direction.}
	\label{fig:boxes_bis}
\end{figure}

We used the extracted scalar field values as a map and plotted the topological charge density within each simplex,
with red for positive and blue for negative charges, as shown in Fig.~\ref{fig:boxes}
for the snapshot of a specific configuration, which is classified in the $Q=1$ topological sector, after 10 cooling steps.
It is interesting to notice that, at this early stage of cooling, the topological charge is still substantially delocalized, i.e., many clusters appear. This is similar to what is observed in standard lattice simulations (see, e.g., Refs.~\cite{Buividovich:2011cv, Ilgenfritz:2007xu, Alexandru:2005bn}).
On the other hand, the same picture after prolongated cooling (see Fig.~\ref{fig:boxes_bis}) appears as a single and well-defined cluster.

Let us now illustrate some more technical details about how the pseudo-Cartesian coordinates are actually implemented.
One way to add matter fields into a triangulated manifold is to treat the field as a map between the triangulated spacetime $\mathcal{M}_H(g_{\mu\nu})$ and a target space $\mathcal{N}(h_{\alpha\beta})$ with fixed metric and topology. In the case where the target space is Euclidean $\mathbb{R}^d$ or $S^d$ with a diagonal and flat metric on $\mathcal{N}$, then the continuum action
\begin{eqnarray}
	S_M[\phi, g]	= \quad \quad \nonumber \\\frac{1}{2} \int \mathrm{d}^{4}x \sqrt{g(x)} \; g^{\mu \nu} (x) \; h_{\rho \sigma}(\phi^\gamma(x)) \nonumber
	\times \partial_\mu \phi^\rho(x) \partial_\nu \phi^\sigma(x) = \\\frac{1}{2} \sum_{\sigma=1}^d  \int \mathrm{d}^{4}x \sqrt{g(x)} \;  \partial^\nu \phi^\sigma(x) \partial_\nu \phi^\sigma(x) \quad \quad \quad
	\label{contmatteraction}
\end{eqnarray}
can be separated for each direction $\sigma$. The addition of a boundary allows the field values to technically obtain a winding number, but the added boundary jump condition (with amplitude $\delta$) renders the mapping to remain continuous by identifying
\begin{equation}\label{identify}
	\phi_i^\sigma \equiv \phi_i^\sigma+n \cdot \delta, \quad n\in \mathbb{Z},
\end{equation}
at the boundary, mapping the field values to a circle with circumference $\delta$. The discrete form of the matter field action with this boundary term is
\begin{eqnarray}
	S_M^{CDT}[\phi,T] = \frac{1}{2} \sum_{\sigma=1}^d \sum_{i \leftrightarrow j}(\phi^\sigma_i-\phi^\sigma_j-\delta \cdot \mathbf{B}^\sigma_{ij})^2 = \nonumber \\
	=\sum_{\sigma=1}^d \left(\sum_{i, j} \phi_i^\sigma \mathbf{L}_{ij}(T) \phi_j^\sigma -2\delta\sum_i \phi_i^\sigma b_i^\sigma + \delta^2\cdot V^\sigma \right)
	\label{actionb}
\end{eqnarray}
where $ \sum_{i \leftrightarrow j}$ counts all of the neighbors of the discrete building blocks and $\mathbf{L}_{ij}(T)$ is defined as the discrete Laplace--Beltrami operator for the underlying dual graph. The boundary term $\mathbf{B}^\sigma_{ij}$ gives a $\pm \delta$ contribution if the particular simplex pair $\{i,j\}$ lies on the boundary. Minimizing the action leads to
\begin{equation}\label{eq:laplace_eq3}
	\mathbf{L}\phi^\sigma =  \delta \cdot b^\sigma,
\end{equation}
which is the inhomogeneous Laplace equation for the scalar field $\phi^\sigma$ with a boundary term $b^\sigma$ with scale $\delta$. Decomposing the scalar field as $\phi^\sigma = \bar \phi^\sigma+ \xi^\sigma$ and expanding the action leads to the form
\begin{eqnarray}
	S_M^{CDT}[\phi,\xi,T] = \sum_{\sigma,i,j}  \xi_i^\sigma \mathbf{L}_{ij}(T) \xi_j^\sigma + \quad \mbox{~~~~~~~~}
	\nonumber \\
	+\sum_{\sigma=1}^d \left(\sum_{i, j}\bar \phi_i^\sigma \mathbf{L}_{ij}(T) \bar\phi_j^\sigma -2\delta\sum_i \bar\phi_i^\sigma b_i^\sigma + \delta^2\cdot V^\sigma \right) \nonumber  \\
	= \sum_{\sigma,i,j}  \xi_i^\sigma \mathbf{L}_{ij}(T) \xi_j^\sigma + S_M^{CDT}[\bar \phi,T]. \quad\mbox{~~~~~~~~}
	\label{actionb2}
\end{eqnarray}
The classical solution $\bar \phi^\sigma$ of the scalar fields obtains a winding number 1 and the quantum fields $\xi^\sigma$ take real values with winding number 0. \\

We took thermalized pure gravity geometries with defined boundaries as a technical tool. For fixed geometries (in the quenched approximation), setting the value of $\delta > 0$ and solving the inhomogeneous Laplace equation, Eq.~\eqref{eq:laplace_eq3}, by cooling the dynamical scalar fields, we effectively forced the quantum fields $\xi^\sigma$ to align with the classical solution $\bar{\phi}^\sigma$. Then, extracting the scalar field coordinates for each simplex in directions ($\phi_x, \phi_y, \phi_z$)
gave us a set of spatial coordinates, and using the foliation $t$ as a fourth coordinate and rescaling all coordinate values to 1 we obtained a $[0,1]^4$ box with a density inside, corresponding to the map from the underlying geometry to the flat box. Associating to each simplex the topological charge density, we can visualize it in this representation as the function of the scalar field coordinates, which is analogous to a Laplacian embedding of a graph. It was shown that this map exhibits homogeneous large-scale properties only in the de Sitter phase.

\end{document}